\documentclass[aps,prl,twocolumn,notitlepage,groupedaddress,showpacs]{revtex4-1}

\usepackage{amsmath}    
\usepackage{amsfonts}
\usepackage{bm}
\usepackage{amssymb}
\usepackage{appendix}
\usepackage{graphicx}   
\usepackage{color}      
\usepackage{subfigure}  
\usepackage{comment}

\newcommand{\im}{\textrm{Im}}

\begin{document}

\title{Achieving arbitrary polarization control using complex birefringent meta-materials}
\author{Alexander Cerjan and Shanhui Fan} 
\affiliation{Department of Electrical Engineering, and Ginzton Laboratory, Stanford  University,  Stanford,  California  94305,  USA}

\date{\today}

\begin{abstract}
We demonstrate that the key to realizing arbitrary control over pairs of polarization states of light, i.e.\ transforming an arbitrarily
polarized pair of input states to an arbitrarily polarized pair of output states, is the ability to
generate pairs of states with orthogonal polarizations from non-orthogonal pairs of initial states. We then develop a
new class of non-Hermitian meta-materials, termed \textit{complex birefringent} meta-materials, which are able to
do exactly this. Such materials could facilitate the detection of small polarization changes in
scattering experiments, as well as enable new polarization multiplexing schemes in communications networks.
\end{abstract}

\maketitle

Polarization is one of the fundamental properties of light, and control over the polarization
is paramount in many optical communications and imaging applications. In general,
the effect of propagation through any media on the polarization of an incident electromagnetic signal can be
described as $|\beta \rangle = S(z) |\alpha \rangle$,
where $|\alpha \rangle$ and $|\beta \rangle$ are the input and output polarization states,
respectively, and $S(z)$ is a $2 \times 2$ matrix that depends on the properties of the medium,
as well as the propagation distance $z$. Conventionally,
the polarization of a signal is manipulated through the use of birefringent materials \cite{imai_optical_1985,okoshi_polarization-state_1985,walker_polarization_1990,zhuang_lcd_1999,hao_manipulating_2007,hao_optical_2009,pors_plasmonic_2011}.
For lossless birefringent media, with proper choice of material parameters and propagation distance, 
it is always possible to convert an input polarization $|\alpha_1 \rangle$ to
an arbitrary output polarization $|\beta_1 \rangle$. However, once the
response to $|\alpha_1 \rangle$ is determined, the output polarization 
$|\beta_2 \rangle = S |\alpha_2 \rangle$ is no longer arbitrary for any other input polarization 
$|\alpha_2 \rangle$. This is because $S$ is unitary in lossless media, and thus
$\langle \beta_2 | \beta_1 \rangle = \langle \alpha_2 | \alpha_1 \rangle$. 


In this Letter, we seek to overcome the limitation of conventional birefringent media by 
developing a class of meta-materials which enable arbitrary control over pairs of polarization states.
By arbitrary control, we demand that for a \textit{pair} of arbitrary input
polarizations $|\alpha_1 \rangle$ and $|\alpha_2 \rangle$, one can generate an arbitrary \textit{pair}
of output polarizations $|\beta_1 \rangle$ and $|\beta_2 \rangle$.
Achieving such polarization control has significant implications for a wide range of technologies.
For example, with this capability one can map
two polarizations that are close to each other into two orthogonal polarizations, which may
facilitate the detection of small polarization changes, such as those arising from the imaging
of biological tissues \cite{de_boer_review_2002,adato_ultra-sensitive_2009} and thin films \cite{losurdo_spectroscopic_2009}. Likewise, the ability to completely
separate non-orthogonal polarization states could enable new multiplexing schemes in 
optical communications networks beyond what is currently possible \cite{hayee_doubling_2001,morant_polarization_2014}.

We first show that the key step for achieving arbitrary control over pairs of polarization states is to develop a
class of meta-materials which are capable of performing the following polarization transformation as
denoted by $S_\theta$,
\begin{align}
&|1,1 \rangle = S_\theta |\theta \rangle, \label{eq:s1} \\
&|1,-1 \rangle = S_\theta |-\theta \rangle. \label{eq:s2}
\end{align}
Here we assume propagation along the $z$-axis, and label the polarization states in terms of the electric
field components in the $xy$-plane as $|E_x, E_y \rangle$. $|\pm \theta \rangle$ denote the two polarization 
states that lie on the great circle of the Poincar\'{e} sphere passing through $|1,i\rangle$ and $|1,1\rangle$, 
and are symmetrically placed away from $|1,i\rangle$, subtending an angle of $\pm \theta$ with respect to $|1, i\rangle$.
Suppose we can construct a class of materials which can provide $S_\theta$ for an arbitrary $\theta$. For an
arbitrary pair of input states $|\alpha_1 \rangle$ and $|\alpha_2 \rangle$, using conventional lossless birefringent materials, one can
achieve the transformation \cite{zhuang_lcd_1999}
\begin{align}
&|\theta_\alpha \rangle = U_\alpha |\alpha_1 \rangle, \\
&|-\theta_\alpha \rangle = U_\alpha |\alpha_2 \rangle,
\end{align}
where $U_\alpha$ is unitary and $\langle \theta_\alpha | -\theta_\alpha \rangle = \langle \alpha_1 | \alpha_2 \rangle$.
For the pair of arbitrary output states $|\beta_1 \rangle$ and
$|\beta_2 \rangle$, one can obtain a similar unitary transformation $U_\beta$ that transforms them to $|\pm \theta_\beta \rangle$.
Therefore, the transformation $S$ from the input states $|\alpha_1 \rangle$ and $|\alpha_2 \rangle$ to the output states $|\beta_1 \rangle$ and
$|\beta_2 \rangle$ is then
\begin{equation}
S = U_\beta^\dagger S_{\theta_\beta}^{-1} S_{\theta_\alpha} U_\alpha.
\end{equation}

To achieve the transformation as described by Eqs.\ (\ref{eq:s1}) and (\ref{eq:s2}) in general requires 
a non-Hermitian meta-material. We now proceed to show that such a transformation can be realized
in a class of complex symmetric meta-materials with its dielectric tensor having the form
\begin{equation}
  \bar{\varepsilon} = \left( 
  \begin{array}{ccc}
    \varepsilon_{xx} & \varepsilon_{xy} & 0 \\
    \varepsilon_{yx} & \varepsilon_{yy} & 0 \\
    0 & 0 & \varepsilon_{zz}
  \end{array}
  \right) 
  \equiv \left( 
  \begin{array}{cc}
  \bar{\varepsilon}_\perp & 0 \\
   0 & \varepsilon_{zz}
  \end{array}
  \right)
  , \label{eq:dielTen}
\end{equation}
in which
\begin{align}
\varepsilon_{xx} &= \varepsilon_{yy}^* = \varepsilon_r - i \varepsilon_i, \label{eq:spec1} \\
\varepsilon_{xy} &= \varepsilon_{yx},  \label{eq:spec3}
\end{align}
where $\varepsilon_r, \varepsilon_i, \varepsilon_{xy} \in \mathbb{R}$. 
Here, an equal amount of gain and
loss has been added to the $x$ and $y$ axes of a conventional birefringent material,
a choice inspired by recent developments in
optical media with spatially distributed regions containing equal amounts of gain and loss \cite{bender_pt-symmetric_1999,bender_complex_2002,musslimani_optical_2008,makris_beam_2008,guo_observation_2009,klaiman_visualization_2008,longhi_bloch_2009,ruter_observation_2010,chong_pt-symmetry_2011,szameit_pt-symmetry_2011,feng_experimental_2013,peng_parity-time-symmetric_2014,chang_parity-time_2014,lawrence_manifestation_2014,kang_coherent_2015,cerjan_zipping_2016}.
Henceforth, we refer to materials which obey Eqs.\ (\ref{eq:spec1}) and (\ref{eq:spec3}) as 
\textit{complex birefringent} meta-materials.

\begin{figure}[t!]
    \centering
    \includegraphics[width=0.48\textwidth]{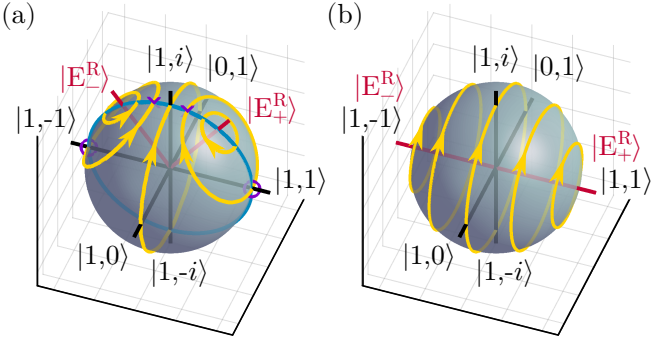}
    \caption{
      (a) Flow lines (yellow) depict the polarization dynamics on the Poincar\'{e} sphere when light travels along the $z$-direction
      through a complex birefringent material as described by Eqs.\ (\ref{eq:dielTen})-(\ref{eq:spec3}) with $\tau = 0.81$. Note that the shape of the flow lines depends only the value of $\tau$. 
      The great circle containing $|1,i\rangle$ 
      and $|1,1\rangle$ is shown in blue. The non-orthogonal eigen-polarizations of the dielectric tensor
      which form the axis of polarization flow are shown in red. The purple crosses 
      show the initial non-orthogonal polarization states $|\pm \theta \rangle$
      which can be mapped to the orthogonal polarization states $|1, \pm 1\rangle$, shown as purple circles,
      as described by Eqs.\ (\ref{eq:s1}) and (\ref{eq:s2}).
      (b) Flow lines (yellow) depict the polarization dynamics on the Poincar\'{e} sphere when traveling
      through a conventional birefringent material, with $\tau = 0$. The orthogonal eigen-polarizations
      of the dielectric tensor are shown in red.
      \label{fig:unbroken}}
\end{figure}

For light propagating along the $z$-axis of such a medium, 
the allowed wavevectors, $k_\pm$, of a monochromatic signal with frequency $\omega$,
can be found by solving the right eigenvalue equation \cite{yeh,haus}, 
\begin{equation}
\omega^2 \mu \bar{\varepsilon}_\perp |E_\pm^R \rangle = k^2 |E_\pm^R \rangle, \label{eq:wave}
\end{equation}
in which $\mu$ is the scalar magnetic permeability.
As such, the allowed wavevectors and right eigen-polarizations of complex birefringent meta-materials are
\begin{align}
\frac{k_{\pm}^2}{\omega^2 \mu} &= \varepsilon_r \pm \varepsilon_{xy} \sqrt{1 - \tau^2}, \label{eq:kpm}\\
|E_\pm^R \rangle &= \frac{1}{N_\pm} \left| 1, i\tau \pm \sqrt{1 - \tau^2} \right\rangle, \label{eq:epm}
\end{align}
in which $\tau = \varepsilon_i / \varepsilon_{xy}$ represents a normalized measure of the strength of the gain and loss in the system,
and
\begin{equation}
N_\pm^2 = 2(1-\tau^2) \pm 2i\tau \sqrt{1-\tau^2}, \label{eq:Npm}
\end{equation}
is the normalization of the eigenstates.
The matrix $\omega^2 \mu \bar{\varepsilon}_\perp$ also has left eigen-polarizations, which are 
solutions to $\langle E_\pm^L | \omega^2 \mu \bar{\varepsilon}_\perp = k_\pm^2 \langle E_\pm^L|$.
Together, the
left and right eigen-polarizations form a bi-orthogonal basis, and can be normalized such that $\langle E_m^L | E_n^R \rangle = \delta_{mn}$, with the choice of $N_\pm$ in Eq.\ (\ref{eq:Npm}).
In conventional lossless birefringent media 
$\bar{\varepsilon}_\perp$ is Hermitian, and so $\langle E_\pm^L | = |E_\pm^R \rangle^\dagger$.
However, for complex birefringent materials, $\omega^2 \mu \bar{\varepsilon}_\perp$ is complex-symmetric, and the left and right
eigen-polarizations are related by $\langle E_\pm^L| = |E_\pm^R \rangle^T$. 
Moreover, although the two right eigen-polarizations are linearly independent for complex birefringent materials with $|\tau| \ne 1$, 
it can be readily seen that they are not orthogonal, $\langle E_\mp^R | E_\pm^R \rangle \ne 0$ except for when $\tau = 0$ and the
system reverts to a conventional birefringent material, or when $|\tau| \rightarrow \infty$ and the
system becomes a conventional dichroic material.

The evolution of the polarization of light propagating within a complex birefringent material with $|\tau| \ne 1$ can
be expressed in terms of the right eigen-polarizations (\ref{eq:epm}), as
\begin{equation}
|E(z)\rangle = e^{ik_- z}\left(e^{i \Delta k z} A_+ |E_+^R \rangle + A_- |E_-^R \rangle \right), \label{eq:field}
\end{equation}
in which $\Delta k = k_+ - k_-$ is the additional phase accumulated by $|E_+^R\rangle$ relative to $|E_-^R\rangle$ per
unit length, and the initial amplitudes, $A_\pm$, are defined in terms of the left eigen-polarizations
as $A_\pm = \langle E_\pm^L | E(0) \rangle$.
The resulting polarization dynamics of a complex birefringent material can be
visualized by plotting the output polarization as a function of $z$ on the Poincar\'{e} sphere. 
The example shown in Fig.\ \ref{fig:unbroken}(a) illustrates the polarization dynamics when $\tau < 1$, 
for which both allowed wavevectors are real, $k_\pm \in \mathbb{R}$. This is analogous to the exact
phase in parity-time symmetric systems \cite{musslimani_optical_2008,makris_beam_2008,guo_observation_2009}. 
For the sake of comparison, we also plot the polarization
dynamics for a conventional Hermitian birefringent material in Fig.\ \ref{fig:unbroken}(b), as described by setting $\varepsilon_i = 0$ 
in Eq.\ (\ref{eq:dielTen}). 
For both types of materials as shown in Fig.\ \ref{fig:unbroken}, the
eigen-polarizations define the fixed point of the dynamics. So long as the incident polarization is not
parallel to one of these eigen-polarizations, the polarization of the initial signal forms a closed
trajectory around the eigen-polarizations as $z$ is varied. 

When $\tau = 0$, which describes a conventional lossless birefrigent material, the two eigen-polarizations
are located at $|1,1\rangle$ and $|1, -1\rangle$, corresponding to two linearly polarized states. The two
eigenstates are orthogonal to each other, and the polarization trajectories form circles around the axis formed by the 
two eigenstates, shown in Fig.\ \ref{fig:unbroken}(b). (On the Poincare sphere, orthogonal states are
represented by antipodal points.) As $\tau$ is increased, so that $0 < \tau < 1$, the two eigen-polarizations
remain on the great circle connecting $|1, 1 \rangle$ and $|1, i \rangle$, but are tilted away from 
the states $|1,\pm 1\rangle$ towards $|1,i\rangle$, which is a manifestation of the non-orthogonality of
these two eigenstates. For $-1 < \tau < 0$, the eigenstates instead tilt away from $|1, \pm 1 \rangle$ towards
$|1,-i \rangle$. As $k_\pm \in \mathbb{R}$, the polarization trajectories are still closed, but are no longer
centered on the axis formed by the eigenstates, as shown in Fig.\ \ref{fig:unbroken}(a).

Examining the polarization dynamics of Fig.\ \ref{fig:unbroken}(a), we note that there are two states lying on the great
circle connecting $|1,1 \rangle$ and $|1, i\rangle$, indicated by the purple crosses, which can be
transformed to the two states $|1,1\rangle$ and $|1, -1 \rangle$, indicated by purple circles, with a proper choice of the propagation distance $l$. 
Therefore, complex birefrigent meta-materials with $0 < |\tau| < 1$ indeed provide the key 
non-trivial step required for achieving arbitrary control over pairs of polarization states, which is to realize $S_\theta$ as defined in Eqs.\ (\ref{eq:s1}) and (\ref{eq:s2}).
Mathematically, for a given pair of input states $|\pm \theta \rangle$, the complex birefringent meta-material
which transforms these two states to $|1, \pm 1 \rangle$ satisfies
\begin{equation}
e^{i \Delta k l} = \frac{\langle E_-^L | \pm \theta \rangle \langle E_+^L | 1, \pm 1\rangle}{\langle E_+^L | \pm \theta \rangle \langle E_-^L | 1, \pm 1\rangle}. \label{eq:mapper}
\end{equation}
Here, as $\tau$ appears in both $\Delta k$ and the left eigenstates $\langle E_\pm^L|$, Eq.\ (\ref{eq:mapper})
represents a complex transcendental equation which can be solved for $\tau$ and $l$.
One can show that for any given value of $0<|\tau|<1$, one can achieve orthogonal output
states for any choice of $\theta$ with the proper choice of propagation distance $l$. In general,
as the choice of initial states become parallel ($\theta \rightarrow 0$), stronger gain
and loss ($|\tau| \rightarrow 1$) along with longer propagation distances are required to
achieve orthogonal output states.

\begin{figure}[tb]
    \centering
    \includegraphics[width=0.48\textwidth]{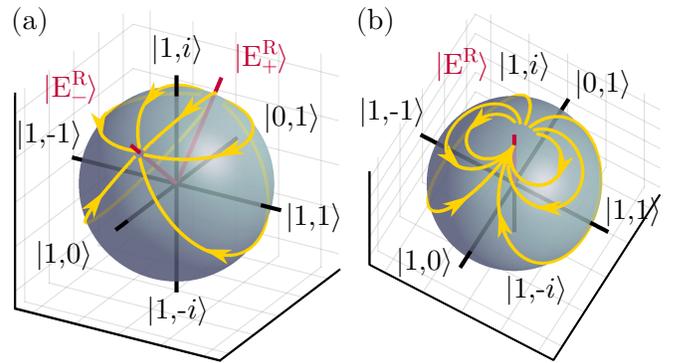}
    \caption{(a) Flow lines (yellow) depict the polarization dynamics on the Poincar\'{e} sphere when light travels along the $z$-direction
      through a complex birefringent material as described by Eqs.\ (\ref{eq:dielTen})-(\ref{eq:spec3}) with $\tau = 1.5$. The eigenvectors of the dielectric tensor
      are shown in red, with the arrow indicating the direction in which the polarization
      flows for $\varepsilon_{xy} > 0$. (b) Flow lines (yellow) depict the polarization dynamics when light travels along the $z$-direction
      through a complex birefringent material as described by Eqs.\ (\ref{eq:dielTen})-(\ref{eq:spec3}) with $\tau = 1$. The single eigen-polarization of the system
      is $|1, i \rangle$ (red).
      \label{fig:broken}}
\end{figure}

Up to this point we have shown that, in the regime where $|\tau|<1$, complex birefringent materials,
as described in Eqs.\ (\ref{eq:dielTen})-(\ref{eq:spec3}), exhibit polarization dynamics that can be used to achieve 
arbitrary control over pairs of polarization states. This class of metamaterials also exhibits interesting polarization dynamics with $|\tau|\ge 1$.
When $|\tau|>1$, both eigenvalues $k_\pm$ become complex, with $k_+ = k_-^*$. 
This is analogous to the broken phase in parity-time symmetric systems.
In this regime the eigen-polarizations reside along the great circle on the Poincar\'{e} sphere 
connecting $|1,0\rangle$, $|0,1 \rangle$, and $|1, \pm i \rangle$.
As $z$ varies, the eigenstates correspond to a stable or an unstable fixed point on the Poincar\'{e} sphere
depending on the sign of $\im[k_\pm]$, as shown in Fig.\ \ref{fig:broken}(a), 
and the material provides polarization-dependent attenuation and
amplification. As $|\tau| \rightarrow \infty$, the system becomes a conventional dichroic material,
with orthogonal eigen-polarizations $|1, 0 \rangle$ and $|0, 1 \rangle$.

When $|\tau|=1$, complex birefringent materials possess an exceptional point \cite{kato,heiss_exceptional_2004}
where both the eigenvalues and eigenvectors coalesce, and the eigenvectors become self-orthogonal, $\langle E_\pm^L | E_\pm^R \rangle = 0$.
This yields two unique properties. First, the expression for the evolution of the electric 
field (\ref{eq:field}) is no longer valid as $\omega^2 \mu \bar{\varepsilon}_\perp$ has a non-trivial
Jordan normal form. Instead, the evolution of the field must be expressed in terms 
of the single remaining eigenvector, $|E^R \rangle$ and its associated Jordan vector, $|J^R \rangle$ \cite{mailybaev,pick_general_2016}, as
\begin{equation}
|E(z)\rangle = e^{ik_0 z}\left[ \left(A_E+iA_J k_0 z\right) |E^R \rangle + A_J |J^R \rangle \right], \label{eq:EPfield}
\end{equation}
in which $k_0 = \omega \sqrt{\mu \varepsilon_r}$, and $A_E,A_J$ are the modal amplitudes at $z=0$.
The derivation of this equation is provided in the Supplementary Information \cite{SI}. 
As can be seen in Eq.\ (\ref{eq:EPfield}), the polarization of light flowing through a complex
birefringent material at $|\tau|=1$ has only a single fixed point that corresponds to circularly polarized light, 
to which the polarization of every initial state with $A_J \ne 0$ 
converges to through \textit{linear} amplification as a function of $z$ and without any attenuation.
Second, any initial polarization
state converges to the fixed point from a single direction, along the $|1,0 \rangle$ to $|1,i \rangle$
contour for $\tau = 1$ as shown in Fig.\ \ref{fig:broken}(b).
This is distinct from what is observed for the stable fixed point when $|\tau|>1$,
in which $|E_-^R\rangle$ can be approached from any direction.

In all three of their phases with $\tau \ne 0$, complex birefringent materials are necessarily 
active optical structures, and in general do not conserve the intensity of the incident radiation.
This fact is immediately evident from the linear and exponential amplification present when $|\tau| \ge 1$.
When $|\tau|<1$, $k_\pm$ are real, and the intensity is instead periodic as a function of $z$,
\begin{align}
I(z) =& |A_+|^2 \langle E_+^R | E_+^R \rangle + A_+^* A_- e^{-i\Delta k z} \langle E_+^R | E_-^R \rangle \notag \\
&+ |A_-|^2 \langle E_-^R | E_-^R \rangle + A_-^* A_+ e^{i\Delta k z} \langle E_-^R | E_+^R \rangle.
\end{align}
Thus, in this case the total change in the intensity is bounded.
Even though the intensity is not conserved in complex birefringent materials, two generalized unitarity relations
can be derived which relate the transmission and reflection coefficients of the scattering matrix, $S$ \cite{chong_pt-symmetry_2011,ge_conservation_2012},
as shown in the Supplementary Information \cite{SI}.
These conserved quantities stem from the fact that complex birefringent materials are invariant upon the operation that switches the
$x$ and $y$ axes of the system, $\mathcal{M}$, and the time-reversal operation, $\mathcal{T}$, thus
\begin{equation}
(\mathcal{MT}) S (\mathcal{MT}) = S^{-1}.
\end{equation}

\begin{figure}[t!]
    \centering
    \subfigure{
      \centering
      \includegraphics[width=0.42\textwidth]{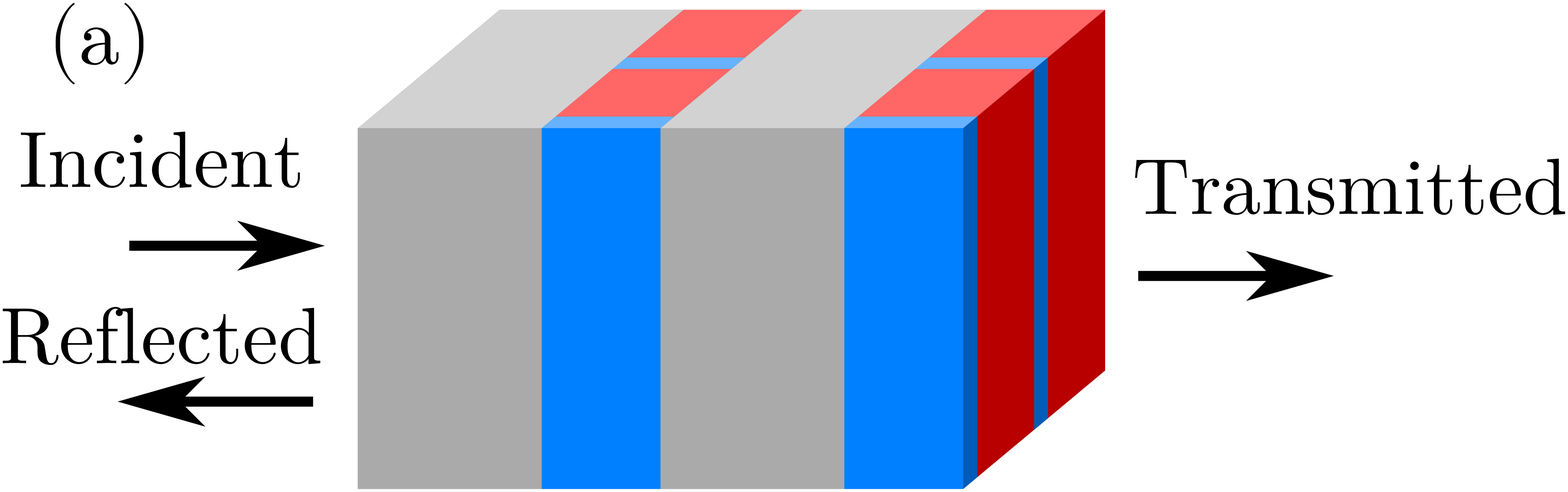}
    }
    \subfigure{
      \centering
      \includegraphics[width=0.25\textwidth]{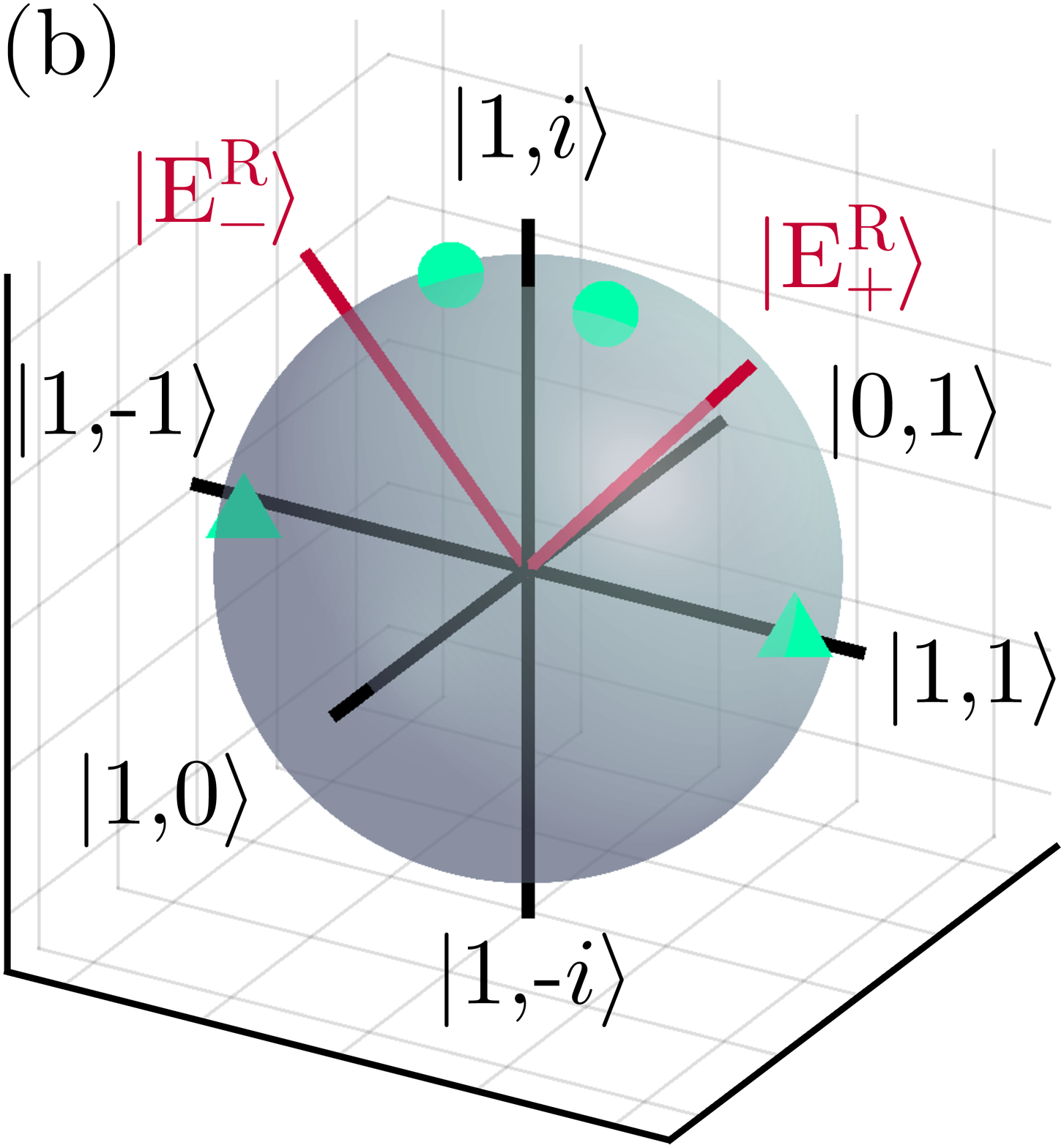}
    }
    \caption{(a) Schematic of a complex birefringent meta-material, consisting of layers of 
      a conventional birefringent material (gray), and layers of a material containing gain (red) and loss (blue), forming a comb.
      The patterning of the structure is assumed to be fine enough relative to the wavelength of the light
      so as to be in the effective medium limit.
      (b) Transformation of the polarization on the Poincar\'{e} sphere
      of two signals through $59 \mu m$ of a complex birefringent meta-material, as shown schematically in (a). 
      Here, for
      an incident light with wavelength $1.55 \mu m$, we have used
      calcium carbonate whose fast and slow axes are rotated $7.47^\circ$ with respect to the lab frame, yielding a dielectric tensor with
      $\varepsilon_{xx} = 2.66$, $\varepsilon_{yy} = 2.19$, $\varepsilon_a = 0.063$. The isotropic gain has $\varepsilon = 4 - 0.1i$,
      and the isotropic loss has $\varepsilon = 9.74 + 0.63i$. By forming a comb consisting of $77\%$ gain regions and $23\%$ loss regions, the effective dielectric tensor
      is anisotropic in this layer, with $\varepsilon_\perp = 5.34 + 0.07i$ and $\varepsilon_\parallel = 4.64 - 0.07i$.
      By using $30nm$ layers of calcium carbonate, and $20nm$ layers of the gain and loss, the total system constitutes a complex birefringent
      metamaterial with $\varepsilon_{xx} = 3.454 - 0.028i$, $\varepsilon_{yy} = 3.453 + 0.028i$, and $\varepsilon_{xy} = 0.038$, which
      corresponds to $\tau = 0.75$.
      The initial signal polarizations are separated by $16.2^\circ$ (cyan circles), 
      while the final polarization states are nearly orthogonal (cyan triangles). The surrounding medium is index matched to the complex birefringent
      meta-material, with $\varepsilon = 3.46$. 
      \label{fig:meta}}
\end{figure}

There are many possible experimental realizations of complex birefringent materials.
A dielectric response as described by Eqs.\ (\ref{eq:dielTen})-(\ref{eq:spec3}) has been previously realized experimentally in a meta-surface 
structure \cite{lawrence_manifestation_2014}. However, in order to observe the polarization dynamics effects, and to achieve the capability for arbitrary 
control over pairs of polarization states, neither of which are considered in \cite{lawrence_manifestation_2014}, it would be interesting to create three-dimensional 
media where the propagation distance can be varied. Therefore, here instead we discuss the construction of 
three-dimensional systems with the appropriate dielectric tensor. As an example, one could construct
a meta-material consisting of an ordinary birefringent material interspersed with layers containing 
regions of both gain and loss, as depicted in Fig.\ \ref{fig:meta}(a).
Transfer matrix calculations of this exact structure without using the effective medium approximation in the
propagation direction confirm its ability to separate
a pair of initial states with similar polarizations to be nearly orthogonal, shown in Fig.\ \ref{fig:meta}(b).
Alternatively, there are many methods for adding birefringence to optical fibers geometrically, allowing for
$\varepsilon_{xy} \ne 0$.
By doping such a birefringent fiber, gain could be added to both $\varepsilon_{xx}$ and $\varepsilon_{yy}$. Then,
all that is required to realize complex birefringence is the ability to add loss specifically to
$\varepsilon_{yy}$.
Regardless of the specific realization chosen, the experimental design of complex birefringent materials
benefits from the critical feature that the amount of gain and loss, $\varepsilon_i$, necessary to observe significant non-trivial polarization dynamics
is set by the anisotropy of the system, $\varepsilon_{xy}$, which can be designed to be quite small. Thus, very little gain or loss
is necessary to realize arbitrary control over pairs of polarization states in these materials. 

Here, we have focused on meta-materials with $\varepsilon_{xx} = \varepsilon_{yy}^*$ as this choice yields
a regime of parameter space, $|\tau| < 1$, where the eigenvalues of the system are real, and as such
the change in intensity of an incident signal is bounded. However,
many other choices of $\varepsilon_{xx},\varepsilon_{yy} \in \mathbb{C}$, such as birefringent materials with loss in a single
polarization channel, i.e.\ $\varepsilon_{xx} = \varepsilon_r$ and $\varepsilon_{yy} = \varepsilon_r + 2i \varepsilon_i$ with $\varepsilon_{xy} \ne 0$, will still yield non-orthogonal
eigenvectors, which can allow for nearly arbitrary control over pairs of polarization states as shown in Fig.\ S2 in the Supplementary Information \cite{SI}.
Similarly,
off-axis propagation in directions which do not conserve $\mathcal{M}$ 
also results in an entirely complex spectrum even for $|\tau|<1$. Fortunately, the rate at which
$k_\pm$ acquires an imaginary component for off-axis propagation is slow relative to the change in propagation angle, 
as shown in Fig.\ S3, and thus the off-axis
components of a wave-packet traveling through a complex birefringent meta-material will 
experience similar polarization dynamics to the on-axis component. 

In conclusion, we have developed a theory of complex symmetric anisotropic dielectric materials and demonstrated
that such systems enable arbitrary control over pairs of polarization states. In particular, such
complex birefringent materials may have applications in both splitting signals with adjacent polarizations
and nearly combining signals with orthogonal polarizations. 
Experimentally, these materials benefit from
the fact that the scale of the gain and loss required to observe these effects is set by
the off-diagonal anisotropy in the system.

\begin{acknowledgments}
We would like to thank Brandon Redding for stimulating discussions about birefringent optical fibers.
This work was supported by an AFOSR MURI program (Grant No.\ FA9550-12-1-0471), and an AFOSR project
(Grant No.\ FA9550-16-1-0010).
\end{acknowledgments}


\begin{thebibliography}{36}%
\makeatletter
\providecommand \@ifxundefined [1]{%
 \@ifx{#1\undefined}
}%
\providecommand \@ifnum [1]{%
 \ifnum #1\expandafter \@firstoftwo
 \else \expandafter \@secondoftwo
 \fi
}%
\providecommand \@ifx [1]{%
 \ifx #1\expandafter \@firstoftwo
 \else \expandafter \@secondoftwo
 \fi
}%
\providecommand \natexlab [1]{#1}%
\providecommand \enquote  [1]{``#1''}%
\providecommand \bibnamefont  [1]{#1}%
\providecommand \bibfnamefont [1]{#1}%
\providecommand \citenamefont [1]{#1}%
\providecommand \href@noop [0]{\@secondoftwo}%
\providecommand \href [0]{\begingroup \@sanitize@url \@href}%
\providecommand \@href[1]{\@@startlink{#1}\@@href}%
\providecommand \@@href[1]{\endgroup#1\@@endlink}%
\providecommand \@sanitize@url [0]{\catcode `\\12\catcode `\$12\catcode
  `\&12\catcode `\#12\catcode `\^12\catcode `\_12\catcode `\%12\relax}%
\providecommand \@@startlink[1]{}%
\providecommand \@@endlink[0]{}%
\providecommand \url  [0]{\begingroup\@sanitize@url \@url }%
\providecommand \@url [1]{\endgroup\@href {#1}{\urlprefix }}%
\providecommand \urlprefix  [0]{URL }%
\providecommand \Eprint [0]{\href }%
\providecommand \doibase [0]{http://dx.doi.org/}%
\providecommand \selectlanguage [0]{\@gobble}%
\providecommand \bibinfo  [0]{\@secondoftwo}%
\providecommand \bibfield  [0]{\@secondoftwo}%
\providecommand \translation [1]{[#1]}%
\providecommand \BibitemOpen [0]{}%
\providecommand \bibitemStop [0]{}%
\providecommand \bibitemNoStop [0]{.\EOS\space}%
\providecommand \EOS [0]{\spacefactor3000\relax}%
\providecommand \BibitemShut  [1]{\csname bibitem#1\endcsname}%
\let\auto@bib@innerbib\@empty
\bibitem [{\citenamefont {Imai}\ \emph {et~al.}(1985)\citenamefont {Imai},
  \citenamefont {Nosu},\ and\ \citenamefont {Yamaguchi}}]{imai_optical_1985}%
  \BibitemOpen
  \bibfield  {author} {\bibinfo {author} {\bibfnamefont {T.}~\bibnamefont
  {Imai}}, \bibinfo {author} {\bibfnamefont {K.}~\bibnamefont {Nosu}}, \ and\
  \bibinfo {author} {\bibfnamefont {H.}~\bibnamefont {Yamaguchi}},\ }\href
  {\doibase 10.1049/el:19850036} {\bibfield  {journal} {\bibinfo  {journal}
  {Electron. Lett.}\ }\textbf {\bibinfo {volume} {21}},\ \bibinfo {pages} {52}
  (\bibinfo {year} {1985})}\BibitemShut {NoStop}%
\bibitem [{\citenamefont {Okoshi}(1985)}]{okoshi_polarization-state_1985}%
  \BibitemOpen
  \bibfield  {author} {\bibinfo {author} {\bibfnamefont {T.}~\bibnamefont
  {Okoshi}},\ }\href {\doibase 10.1109/JLT.1985.1074336} {\bibfield  {journal}
  {\bibinfo  {journal} {J. Lightwave Technol.}\ }\textbf {\bibinfo {volume}
  {3}},\ \bibinfo {pages} {1232} (\bibinfo {year} {1985})}\BibitemShut
  {NoStop}%
\bibitem [{\citenamefont {Walker}\ and\ \citenamefont
  {Walker}(1990)}]{walker_polarization_1990}%
  \BibitemOpen
  \bibfield  {author} {\bibinfo {author} {\bibfnamefont {N.~G.}\ \bibnamefont
  {Walker}}\ and\ \bibinfo {author} {\bibfnamefont {G.~R.}\ \bibnamefont
  {Walker}},\ }\href {\doibase 10.1109/50.50740} {\bibfield  {journal}
  {\bibinfo  {journal} {J. Lightwave Technol.}\ }\textbf {\bibinfo {volume}
  {8}},\ \bibinfo {pages} {438} (\bibinfo {year} {1990})}\BibitemShut {NoStop}%
\bibitem [{\citenamefont {Zhuang}\ \emph {et~al.}(1999)\citenamefont {Zhuang},
  \citenamefont {Suh},\ and\ \citenamefont {Patel}}]{zhuang_lcd_1999}%
  \BibitemOpen
  \bibfield  {author} {\bibinfo {author} {\bibfnamefont {Z.}~\bibnamefont
  {Zhuang}}, \bibinfo {author} {\bibfnamefont {S.-W.}\ \bibnamefont {Suh}}, \
  and\ \bibinfo {author} {\bibfnamefont {J.~S.}\ \bibnamefont {Patel}},\ }\href
  {\doibase 10.1364/OL.24.000694} {\bibfield  {journal} {\bibinfo  {journal}
  {Opt. Lett.}\ }\textbf {\bibinfo {volume} {24}},\ \bibinfo {pages} {694}
  (\bibinfo {year} {1999})}\BibitemShut {NoStop}%
\bibitem [{\citenamefont {Hao}\ \emph {et~al.}(2007)\citenamefont {Hao},
  \citenamefont {Yuan}, \citenamefont {Ran}, \citenamefont {Jiang},
  \citenamefont {Kong}, \citenamefont {Chan},\ and\ \citenamefont
  {Zhou}}]{hao_manipulating_2007}%
  \BibitemOpen
  \bibfield  {author} {\bibinfo {author} {\bibfnamefont {J.}~\bibnamefont
  {Hao}}, \bibinfo {author} {\bibfnamefont {Y.}~\bibnamefont {Yuan}}, \bibinfo
  {author} {\bibfnamefont {L.}~\bibnamefont {Ran}}, \bibinfo {author}
  {\bibfnamefont {T.}~\bibnamefont {Jiang}}, \bibinfo {author} {\bibfnamefont
  {J.~A.}\ \bibnamefont {Kong}}, \bibinfo {author} {\bibfnamefont {C.~T.}\
  \bibnamefont {Chan}}, \ and\ \bibinfo {author} {\bibfnamefont
  {L.}~\bibnamefont {Zhou}},\ }\href {\doibase 10.1103/PhysRevLett.99.063908}
  {\bibfield  {journal} {\bibinfo  {journal} {Phys. Rev. Lett.}\ }\textbf
  {\bibinfo {volume} {99}},\ \bibinfo {pages} {063908} (\bibinfo {year}
  {2007})}\BibitemShut {NoStop}%
\bibitem [{\citenamefont {Hao}\ \emph {et~al.}(2009)\citenamefont {Hao},
  \citenamefont {Ren}, \citenamefont {An}, \citenamefont {Huang}, \citenamefont
  {Chen}, \citenamefont {Qiu},\ and\ \citenamefont {Zhou}}]{hao_optical_2009}%
  \BibitemOpen
  \bibfield  {author} {\bibinfo {author} {\bibfnamefont {J.}~\bibnamefont
  {Hao}}, \bibinfo {author} {\bibfnamefont {Q.}~\bibnamefont {Ren}}, \bibinfo
  {author} {\bibfnamefont {Z.}~\bibnamefont {An}}, \bibinfo {author}
  {\bibfnamefont {X.}~\bibnamefont {Huang}}, \bibinfo {author} {\bibfnamefont
  {Z.}~\bibnamefont {Chen}}, \bibinfo {author} {\bibfnamefont {M.}~\bibnamefont
  {Qiu}}, \ and\ \bibinfo {author} {\bibfnamefont {L.}~\bibnamefont {Zhou}},\
  }\href {\doibase 10.1103/PhysRevA.80.023807} {\bibfield  {journal} {\bibinfo
  {journal} {Phys. Rev. A}\ }\textbf {\bibinfo {volume} {80}},\ \bibinfo
  {pages} {023807} (\bibinfo {year} {2009})}\BibitemShut {NoStop}%
\bibitem [{\citenamefont {Pors}\ \emph {et~al.}(2011)\citenamefont {Pors},
  \citenamefont {Nielsen}, \citenamefont {Valle}, \citenamefont {Willatzen},
  \citenamefont {Albrektsen},\ and\ \citenamefont
  {Bozhevolnyi}}]{pors_plasmonic_2011}%
  \BibitemOpen
  \bibfield  {author} {\bibinfo {author} {\bibfnamefont {A.}~\bibnamefont
  {Pors}}, \bibinfo {author} {\bibfnamefont {M.~G.}\ \bibnamefont {Nielsen}},
  \bibinfo {author} {\bibfnamefont {G.~D.}\ \bibnamefont {Valle}}, \bibinfo
  {author} {\bibfnamefont {M.}~\bibnamefont {Willatzen}}, \bibinfo {author}
  {\bibfnamefont {O.}~\bibnamefont {Albrektsen}}, \ and\ \bibinfo {author}
  {\bibfnamefont {S.~I.}\ \bibnamefont {Bozhevolnyi}},\ }\href {\doibase
  10.1364/OL.36.001626} {\bibfield  {journal} {\bibinfo  {journal} {Opt.
  Lett.}\ }\textbf {\bibinfo {volume} {36}},\ \bibinfo {pages} {1626} (\bibinfo
  {year} {2011})}\BibitemShut {NoStop}%
\bibitem [{\citenamefont {De~Boer}\ and\ \citenamefont
  {Milner}(2002)}]{de_boer_review_2002}%
  \BibitemOpen
  \bibfield  {author} {\bibinfo {author} {\bibfnamefont {J.~F.}\ \bibnamefont
  {De~Boer}}\ and\ \bibinfo {author} {\bibfnamefont {T.~E.}\ \bibnamefont
  {Milner}},\ }\href
  {http://biomedicaloptics.spiedigitallibrary.org/article.aspx?articleid=1101500}
  {\bibfield  {journal} {\bibinfo  {journal} {J. Biomed. Opt.}\ }\textbf
  {\bibinfo {volume} {7}},\ \bibinfo {pages} {359} (\bibinfo {year}
  {2002})}\BibitemShut {NoStop}%
\bibitem [{\citenamefont {Adato}\ \emph {et~al.}(2009)\citenamefont {Adato},
  \citenamefont {Yanik}, \citenamefont {Amsden}, \citenamefont {Kaplan},
  \citenamefont {Omenetto}, \citenamefont {Hong}, \citenamefont {Erramilli},\
  and\ \citenamefont {Altug}}]{adato_ultra-sensitive_2009}%
  \BibitemOpen
  \bibfield  {author} {\bibinfo {author} {\bibfnamefont {R.}~\bibnamefont
  {Adato}}, \bibinfo {author} {\bibfnamefont {A.~A.}\ \bibnamefont {Yanik}},
  \bibinfo {author} {\bibfnamefont {J.~J.}\ \bibnamefont {Amsden}}, \bibinfo
  {author} {\bibfnamefont {D.~L.}\ \bibnamefont {Kaplan}}, \bibinfo {author}
  {\bibfnamefont {F.~G.}\ \bibnamefont {Omenetto}}, \bibinfo {author}
  {\bibfnamefont {M.~K.}\ \bibnamefont {Hong}}, \bibinfo {author}
  {\bibfnamefont {S.}~\bibnamefont {Erramilli}}, \ and\ \bibinfo {author}
  {\bibfnamefont {H.}~\bibnamefont {Altug}},\ }\href {\doibase
  10.1073/pnas.0907459106} {\bibfield  {journal} {\bibinfo  {journal} {Proc.
  Natl. Acad. Sci. U.S.A.}\ }\textbf {\bibinfo {volume} {106}},\ \bibinfo
  {pages} {19227} (\bibinfo {year} {2009})}\BibitemShut {NoStop}%
\bibitem [{\citenamefont {Losurdo}\ \emph {et~al.}(2009)\citenamefont
  {Losurdo}, \citenamefont {Bergmair}, \citenamefont {Bruno}, \citenamefont
  {Cattelan}, \citenamefont {Cobet}, \citenamefont {Martino}, \citenamefont
  {Fleischer}, \citenamefont {Dohcevic-Mitrovic}, \citenamefont {Esser},
  \citenamefont {Galliet}, \citenamefont {Gajic}, \citenamefont {Hemzal},
  \citenamefont {Hingerl}, \citenamefont {Humlicek}, \citenamefont
  {Ossikovski}, \citenamefont {Popovic},\ and\ \citenamefont
  {Saxl}}]{losurdo_spectroscopic_2009}%
  \BibitemOpen
  \bibfield  {author} {\bibinfo {author} {\bibfnamefont {M.}~\bibnamefont
  {Losurdo}}, \bibinfo {author} {\bibfnamefont {M.}~\bibnamefont {Bergmair}},
  \bibinfo {author} {\bibfnamefont {G.}~\bibnamefont {Bruno}}, \bibinfo
  {author} {\bibfnamefont {D.}~\bibnamefont {Cattelan}}, \bibinfo {author}
  {\bibfnamefont {C.}~\bibnamefont {Cobet}}, \bibinfo {author} {\bibfnamefont
  {A.~d.}\ \bibnamefont {Martino}}, \bibinfo {author} {\bibfnamefont
  {K.}~\bibnamefont {Fleischer}}, \bibinfo {author} {\bibfnamefont
  {Z.}~\bibnamefont {Dohcevic-Mitrovic}}, \bibinfo {author} {\bibfnamefont
  {N.}~\bibnamefont {Esser}}, \bibinfo {author} {\bibfnamefont
  {M.}~\bibnamefont {Galliet}}, \bibinfo {author} {\bibfnamefont
  {R.}~\bibnamefont {Gajic}}, \bibinfo {author} {\bibfnamefont
  {D.}~\bibnamefont {Hemzal}}, \bibinfo {author} {\bibfnamefont
  {K.}~\bibnamefont {Hingerl}}, \bibinfo {author} {\bibfnamefont
  {J.}~\bibnamefont {Humlicek}}, \bibinfo {author} {\bibfnamefont
  {R.}~\bibnamefont {Ossikovski}}, \bibinfo {author} {\bibfnamefont {Z.~V.}\
  \bibnamefont {Popovic}}, \ and\ \bibinfo {author} {\bibfnamefont
  {O.}~\bibnamefont {Saxl}},\ }\href {\doibase 10.1007/s11051-009-9662-6}
  {\bibfield  {journal} {\bibinfo  {journal} {J. Nanoparticle Res.}\ }\textbf
  {\bibinfo {volume} {11}},\ \bibinfo {pages} {1521} (\bibinfo {year}
  {2009})}\BibitemShut {NoStop}%
\bibitem [{\citenamefont {Hayee}\ \emph {et~al.}(2001)\citenamefont {Hayee},
  \citenamefont {Cardakli}, \citenamefont {Sahin},\ and\ \citenamefont
  {Willner}}]{hayee_doubling_2001}%
  \BibitemOpen
  \bibfield  {author} {\bibinfo {author} {\bibfnamefont {M.~I.}\ \bibnamefont
  {Hayee}}, \bibinfo {author} {\bibfnamefont {M.~C.}\ \bibnamefont {Cardakli}},
  \bibinfo {author} {\bibfnamefont {A.~B.}\ \bibnamefont {Sahin}}, \ and\
  \bibinfo {author} {\bibfnamefont {A.~E.}\ \bibnamefont {Willner}},\ }\href
  {\doibase 10.1109/68.935835} {\bibfield  {journal} {\bibinfo  {journal} {IEEE
  Photon. Techol. Lett.}\ }\textbf {\bibinfo {volume} {13}},\ \bibinfo {pages}
  {881} (\bibinfo {year} {2001})}\BibitemShut {NoStop}%
\bibitem [{\citenamefont {Morant}\ \emph {et~al.}(2014)\citenamefont {Morant},
  \citenamefont {P{\' e}rez},\ and\ \citenamefont
  {Llorente}}]{morant_polarization_2014}%
  \BibitemOpen
  \bibfield  {author} {\bibinfo {author} {\bibfnamefont {M.}~\bibnamefont
  {Morant}}, \bibinfo {author} {\bibfnamefont {J.}~\bibnamefont {P{\' e}rez}},
  \ and\ \bibinfo {author} {\bibfnamefont {R.}~\bibnamefont {Llorente}},\
  }\href {\doibase 10.1155/2014/269524} {\bibfield  {journal} {\bibinfo
  {journal} {Adv. Opt. Technol.}\ }\textbf {\bibinfo {volume} {2014}},\
  \bibinfo {pages} {e269524} (\bibinfo {year} {2014})}\BibitemShut {NoStop}%
\bibitem [{\citenamefont {Bender}\ \emph {et~al.}(1999)\citenamefont {Bender},
  \citenamefont {Boettcher},\ and\ \citenamefont
  {Meisinger}}]{bender_pt-symmetric_1999}%
  \BibitemOpen
  \bibfield  {author} {\bibinfo {author} {\bibfnamefont {C.~M.}\ \bibnamefont
  {Bender}}, \bibinfo {author} {\bibfnamefont {S.}~\bibnamefont {Boettcher}}, \
  and\ \bibinfo {author} {\bibfnamefont {P.~N.}\ \bibnamefont {Meisinger}},\
  }\href {\doibase 10.1063/1.532860} {\bibfield  {journal} {\bibinfo  {journal}
  {J. Math. Phys.}\ }\textbf {\bibinfo {volume} {40}},\ \bibinfo {pages} {2201}
  (\bibinfo {year} {1999})}\BibitemShut {NoStop}%
\bibitem [{\citenamefont {Bender}\ \emph {et~al.}(2002)\citenamefont {Bender},
  \citenamefont {Brody},\ and\ \citenamefont {Jones}}]{bender_complex_2002}%
  \BibitemOpen
  \bibfield  {author} {\bibinfo {author} {\bibfnamefont {C.~M.}\ \bibnamefont
  {Bender}}, \bibinfo {author} {\bibfnamefont {D.~C.}\ \bibnamefont {Brody}}, \
  and\ \bibinfo {author} {\bibfnamefont {H.~F.}\ \bibnamefont {Jones}},\ }\href
  {\doibase 10.1103/PhysRevLett.89.270401} {\bibfield  {journal} {\bibinfo
  {journal} {Phys. Rev. Lett.}\ }\textbf {\bibinfo {volume} {89}},\ \bibinfo
  {pages} {270401} (\bibinfo {year} {2002})}\BibitemShut {NoStop}%
\bibitem [{\citenamefont {Musslimani}\ \emph {et~al.}(2008)\citenamefont
  {Musslimani}, \citenamefont {Makris}, \citenamefont {El-Ganainy},\ and\
  \citenamefont {Christodoulides}}]{musslimani_optical_2008}%
  \BibitemOpen
  \bibfield  {author} {\bibinfo {author} {\bibfnamefont {Z.~H.}\ \bibnamefont
  {Musslimani}}, \bibinfo {author} {\bibfnamefont {K.~G.}\ \bibnamefont
  {Makris}}, \bibinfo {author} {\bibfnamefont {R.}~\bibnamefont {El-Ganainy}},
  \ and\ \bibinfo {author} {\bibfnamefont {D.~N.}\ \bibnamefont
  {Christodoulides}},\ }\href {\doibase 10.1103/PhysRevLett.100.030402}
  {\bibfield  {journal} {\bibinfo  {journal} {Phys. Rev. Lett.}\ }\textbf
  {\bibinfo {volume} {100}},\ \bibinfo {pages} {030402} (\bibinfo {year}
  {2008})}\BibitemShut {NoStop}%
\bibitem [{\citenamefont {Makris}\ \emph {et~al.}(2008)\citenamefont {Makris},
  \citenamefont {El-Ganainy}, \citenamefont {Christodoulides},\ and\
  \citenamefont {Musslimani}}]{makris_beam_2008}%
  \BibitemOpen
  \bibfield  {author} {\bibinfo {author} {\bibfnamefont {K.~G.}\ \bibnamefont
  {Makris}}, \bibinfo {author} {\bibfnamefont {R.}~\bibnamefont {El-Ganainy}},
  \bibinfo {author} {\bibfnamefont {D.~N.}\ \bibnamefont {Christodoulides}}, \
  and\ \bibinfo {author} {\bibfnamefont {Z.~H.}\ \bibnamefont {Musslimani}},\
  }\href {\doibase 10.1103/PhysRevLett.100.103904} {\bibfield  {journal}
  {\bibinfo  {journal} {Phys. Rev. Lett.}\ }\textbf {\bibinfo {volume} {100}},\
  \bibinfo {pages} {103904} (\bibinfo {year} {2008})}\BibitemShut {NoStop}%
\bibitem [{\citenamefont {Guo}\ \emph {et~al.}(2009)\citenamefont {Guo},
  \citenamefont {Salamo}, \citenamefont {Duchesne}, \citenamefont {Morandotti},
  \citenamefont {Volatier-Ravat}, \citenamefont {Aimez}, \citenamefont
  {Siviloglou},\ and\ \citenamefont {Christodoulides}}]{guo_observation_2009}%
  \BibitemOpen
  \bibfield  {author} {\bibinfo {author} {\bibfnamefont {A.}~\bibnamefont
  {Guo}}, \bibinfo {author} {\bibfnamefont {G.~J.}\ \bibnamefont {Salamo}},
  \bibinfo {author} {\bibfnamefont {D.}~\bibnamefont {Duchesne}}, \bibinfo
  {author} {\bibfnamefont {R.}~\bibnamefont {Morandotti}}, \bibinfo {author}
  {\bibfnamefont {M.}~\bibnamefont {Volatier-Ravat}}, \bibinfo {author}
  {\bibfnamefont {V.}~\bibnamefont {Aimez}}, \bibinfo {author} {\bibfnamefont
  {G.~A.}\ \bibnamefont {Siviloglou}}, \ and\ \bibinfo {author} {\bibfnamefont
  {D.~N.}\ \bibnamefont {Christodoulides}},\ }\href {\doibase
  10.1103/PhysRevLett.103.093902} {\bibfield  {journal} {\bibinfo  {journal}
  {Phys. Rev. Lett.}\ }\textbf {\bibinfo {volume} {103}},\ \bibinfo {pages}
  {093902} (\bibinfo {year} {2009})}\BibitemShut {NoStop}%
\bibitem [{\citenamefont {Klaiman}\ \emph {et~al.}(2008)\citenamefont
  {Klaiman}, \citenamefont {G\"{u}nther},\ and\ \citenamefont
  {Moiseyev}}]{klaiman_visualization_2008}%
  \BibitemOpen
  \bibfield  {author} {\bibinfo {author} {\bibfnamefont {S.}~\bibnamefont
  {Klaiman}}, \bibinfo {author} {\bibfnamefont {U.}~\bibnamefont
  {G\"{u}nther}}, \ and\ \bibinfo {author} {\bibfnamefont {N.}~\bibnamefont
  {Moiseyev}},\ }\href {\doibase 10.1103/PhysRevLett.101.080402} {\bibfield
  {journal} {\bibinfo  {journal} {Phys. Rev. Lett.}\ }\textbf {\bibinfo
  {volume} {101}},\ \bibinfo {pages} {080402} (\bibinfo {year}
  {2008})}\BibitemShut {NoStop}%
\bibitem [{\citenamefont {Longhi}(2009)}]{longhi_bloch_2009}%
  \BibitemOpen
  \bibfield  {author} {\bibinfo {author} {\bibfnamefont {S.}~\bibnamefont
  {Longhi}},\ }\href {\doibase 10.1103/PhysRevLett.103.123601} {\bibfield
  {journal} {\bibinfo  {journal} {Phys. Rev. Lett.}\ }\textbf {\bibinfo
  {volume} {103}},\ \bibinfo {pages} {123601} (\bibinfo {year}
  {2009})}\BibitemShut {NoStop}%
\bibitem [{\citenamefont {R\"{u}ter}\ \emph {et~al.}(2010)\citenamefont
  {R\"{u}ter}, \citenamefont {Makris}, \citenamefont {El-Ganainy},
  \citenamefont {Christodoulides}, \citenamefont {Segev},\ and\ \citenamefont
  {Kip}}]{ruter_observation_2010}%
  \BibitemOpen
  \bibfield  {author} {\bibinfo {author} {\bibfnamefont {C.~E.}\ \bibnamefont
  {R\"{u}ter}}, \bibinfo {author} {\bibfnamefont {K.~G.}\ \bibnamefont
  {Makris}}, \bibinfo {author} {\bibfnamefont {R.}~\bibnamefont {El-Ganainy}},
  \bibinfo {author} {\bibfnamefont {D.~N.}\ \bibnamefont {Christodoulides}},
  \bibinfo {author} {\bibfnamefont {M.}~\bibnamefont {Segev}}, \ and\ \bibinfo
  {author} {\bibfnamefont {D.}~\bibnamefont {Kip}},\ }\href {\doibase
  10.1038/nphys1515} {\bibfield  {journal} {\bibinfo  {journal} {Nat. Phys.}\
  }\textbf {\bibinfo {volume} {6}},\ \bibinfo {pages} {192} (\bibinfo {year}
  {2010})}\BibitemShut {NoStop}%
\bibitem [{\citenamefont {Chong}\ \emph {et~al.}(2011)\citenamefont {Chong},
  \citenamefont {Ge},\ and\ \citenamefont {Stone}}]{chong_pt-symmetry_2011}%
  \BibitemOpen
  \bibfield  {author} {\bibinfo {author} {\bibfnamefont {Y.~D.}\ \bibnamefont
  {Chong}}, \bibinfo {author} {\bibfnamefont {L.}~\bibnamefont {Ge}}, \ and\
  \bibinfo {author} {\bibfnamefont {A.~D.}\ \bibnamefont {Stone}},\ }\href
  {\doibase 10.1103/PhysRevLett.106.093902} {\bibfield  {journal} {\bibinfo
  {journal} {Phys. Rev. Lett.}\ }\textbf {\bibinfo {volume} {106}},\ \bibinfo
  {pages} {093902} (\bibinfo {year} {2011})}\BibitemShut {NoStop}%
\bibitem [{\citenamefont {Szameit}\ \emph {et~al.}(2011)\citenamefont
  {Szameit}, \citenamefont {Rechtsman}, \citenamefont {Bahat-Treidel},\ and\
  \citenamefont {Segev}}]{szameit_pt-symmetry_2011}%
  \BibitemOpen
  \bibfield  {author} {\bibinfo {author} {\bibfnamefont {A.}~\bibnamefont
  {Szameit}}, \bibinfo {author} {\bibfnamefont {M.~C.}\ \bibnamefont
  {Rechtsman}}, \bibinfo {author} {\bibfnamefont {O.}~\bibnamefont
  {Bahat-Treidel}}, \ and\ \bibinfo {author} {\bibfnamefont {M.}~\bibnamefont
  {Segev}},\ }\href {\doibase 10.1103/PhysRevA.84.021806} {\bibfield  {journal}
  {\bibinfo  {journal} {Phys. Rev. A}\ }\textbf {\bibinfo {volume} {84}},\
  \bibinfo {pages} {021806} (\bibinfo {year} {2011})}\BibitemShut {NoStop}%
\bibitem [{\citenamefont {Feng}\ \emph {et~al.}(2013)\citenamefont {Feng},
  \citenamefont {Xu}, \citenamefont {Fegadolli}, \citenamefont {Lu},
  \citenamefont {Oliveira}, \citenamefont {Almeida}, \citenamefont {Chen},\
  and\ \citenamefont {Scherer}}]{feng_experimental_2013}%
  \BibitemOpen
  \bibfield  {author} {\bibinfo {author} {\bibfnamefont {L.}~\bibnamefont
  {Feng}}, \bibinfo {author} {\bibfnamefont {Y.-L.}\ \bibnamefont {Xu}},
  \bibinfo {author} {\bibfnamefont {W.~S.}\ \bibnamefont {Fegadolli}}, \bibinfo
  {author} {\bibfnamefont {M.-H.}\ \bibnamefont {Lu}}, \bibinfo {author}
  {\bibfnamefont {J.~E.~B.}\ \bibnamefont {Oliveira}}, \bibinfo {author}
  {\bibfnamefont {V.~R.}\ \bibnamefont {Almeida}}, \bibinfo {author}
  {\bibfnamefont {Y.-F.}\ \bibnamefont {Chen}}, \ and\ \bibinfo {author}
  {\bibfnamefont {A.}~\bibnamefont {Scherer}},\ }\href {\doibase
  10.1038/nmat3495} {\bibfield  {journal} {\bibinfo  {journal} {Nat. Mater.}\
  }\textbf {\bibinfo {volume} {12}},\ \bibinfo {pages} {108} (\bibinfo {year}
  {2013})}\BibitemShut {NoStop}%
\bibitem [{\citenamefont {Peng}\ \emph {et~al.}(2014)\citenamefont {Peng},
  \citenamefont {{\" O}zdemir}, \citenamefont {Lei}, \citenamefont {Monifi},
  \citenamefont {Gianfreda}, \citenamefont {Long}, \citenamefont {Fan},
  \citenamefont {Nori}, \citenamefont {Bender},\ and\ \citenamefont
  {Yang}}]{peng_parity-time-symmetric_2014}%
  \BibitemOpen
  \bibfield  {author} {\bibinfo {author} {\bibfnamefont {B.}~\bibnamefont
  {Peng}}, \bibinfo {author} {\bibfnamefont {{\c S}.~K.}\ \bibnamefont {{\"
  O}zdemir}}, \bibinfo {author} {\bibfnamefont {F.}~\bibnamefont {Lei}},
  \bibinfo {author} {\bibfnamefont {F.}~\bibnamefont {Monifi}}, \bibinfo
  {author} {\bibfnamefont {M.}~\bibnamefont {Gianfreda}}, \bibinfo {author}
  {\bibfnamefont {G.~L.}\ \bibnamefont {Long}}, \bibinfo {author}
  {\bibfnamefont {S.}~\bibnamefont {Fan}}, \bibinfo {author} {\bibfnamefont
  {F.}~\bibnamefont {Nori}}, \bibinfo {author} {\bibfnamefont {C.~M.}\
  \bibnamefont {Bender}}, \ and\ \bibinfo {author} {\bibfnamefont
  {L.}~\bibnamefont {Yang}},\ }\href {\doibase 10.1038/nphys2927} {\bibfield
  {journal} {\bibinfo  {journal} {Nat. Phys.}\ }\textbf {\bibinfo {volume}
  {10}},\ \bibinfo {pages} {394} (\bibinfo {year} {2014})}\BibitemShut
  {NoStop}%
\bibitem [{\citenamefont {Chang}\ \emph {et~al.}(2014)\citenamefont {Chang},
  \citenamefont {Jiang}, \citenamefont {Hua}, \citenamefont {Yang},
  \citenamefont {Wen}, \citenamefont {Jiang}, \citenamefont {Li}, \citenamefont
  {Wang},\ and\ \citenamefont {Xiao}}]{chang_parity-time_2014}%
  \BibitemOpen
  \bibfield  {author} {\bibinfo {author} {\bibfnamefont {L.}~\bibnamefont
  {Chang}}, \bibinfo {author} {\bibfnamefont {X.}~\bibnamefont {Jiang}},
  \bibinfo {author} {\bibfnamefont {S.}~\bibnamefont {Hua}}, \bibinfo {author}
  {\bibfnamefont {C.}~\bibnamefont {Yang}}, \bibinfo {author} {\bibfnamefont
  {J.}~\bibnamefont {Wen}}, \bibinfo {author} {\bibfnamefont {L.}~\bibnamefont
  {Jiang}}, \bibinfo {author} {\bibfnamefont {G.}~\bibnamefont {Li}}, \bibinfo
  {author} {\bibfnamefont {G.}~\bibnamefont {Wang}}, \ and\ \bibinfo {author}
  {\bibfnamefont {M.}~\bibnamefont {Xiao}},\ }\href {\doibase
  10.1038/nphoton.2014.133} {\bibfield  {journal} {\bibinfo  {journal} {Nat.
  Photonics}\ }\textbf {\bibinfo {volume} {8}},\ \bibinfo {pages} {524}
  (\bibinfo {year} {2014})}\BibitemShut {NoStop}%
\bibitem [{\citenamefont {Lawrence}\ \emph {et~al.}(2014)\citenamefont
  {Lawrence}, \citenamefont {Xu}, \citenamefont {Zhang}, \citenamefont {Cong},
  \citenamefont {Han}, \citenamefont {Zhang},\ and\ \citenamefont
  {Zhang}}]{lawrence_manifestation_2014}%
  \BibitemOpen
  \bibfield  {author} {\bibinfo {author} {\bibfnamefont {M.}~\bibnamefont
  {Lawrence}}, \bibinfo {author} {\bibfnamefont {N.}~\bibnamefont {Xu}},
  \bibinfo {author} {\bibfnamefont {X.}~\bibnamefont {Zhang}}, \bibinfo
  {author} {\bibfnamefont {L.}~\bibnamefont {Cong}}, \bibinfo {author}
  {\bibfnamefont {J.}~\bibnamefont {Han}}, \bibinfo {author} {\bibfnamefont
  {W.}~\bibnamefont {Zhang}}, \ and\ \bibinfo {author} {\bibfnamefont
  {S.}~\bibnamefont {Zhang}},\ }\href {\doibase 10.1103/PhysRevLett.113.093901}
  {\bibfield  {journal} {\bibinfo  {journal} {Phys. Rev. Lett.}\ }\textbf
  {\bibinfo {volume} {113}},\ \bibinfo {pages} {093901} (\bibinfo {year}
  {2014})}\BibitemShut {NoStop}%
\bibitem [{\citenamefont {Kang}\ and\ \citenamefont
  {Chong}(2015)}]{kang_coherent_2015}%
  \BibitemOpen
  \bibfield  {author} {\bibinfo {author} {\bibfnamefont {M.}~\bibnamefont
  {Kang}}\ and\ \bibinfo {author} {\bibfnamefont {Y.~D.}\ \bibnamefont
  {Chong}},\ }\href {\doibase 10.1103/PhysRevA.92.043826} {\bibfield  {journal}
  {\bibinfo  {journal} {Phys. Rev. A}\ }\textbf {\bibinfo {volume} {92}},\
  \bibinfo {pages} {043826} (\bibinfo {year} {2015})}\BibitemShut {NoStop}%
\bibitem [{\citenamefont {Cerjan}\ \emph {et~al.}(2016)\citenamefont {Cerjan},
  \citenamefont {Raman},\ and\ \citenamefont {Fan}}]{cerjan_zipping_2016}%
  \BibitemOpen
  \bibfield  {author} {\bibinfo {author} {\bibfnamefont {A.}~\bibnamefont
  {Cerjan}}, \bibinfo {author} {\bibfnamefont {A.}~\bibnamefont {Raman}}, \
  and\ \bibinfo {author} {\bibfnamefont {S.}~\bibnamefont {Fan}},\ }\href
  {\doibase 10.1103/PhysRevLett.116.203902} {\bibfield  {journal} {\bibinfo
  {journal} {Phys. Rev. Lett.}\ }\textbf {\bibinfo {volume} {116}},\ \bibinfo
  {pages} {203902} (\bibinfo {year} {2016})}\BibitemShut {NoStop}%
\bibitem [{\citenamefont {Yeh}(2005)}]{yeh}%
  \BibitemOpen
  \bibfield  {author} {\bibinfo {author} {\bibfnamefont {P.}~\bibnamefont
  {Yeh}},\ }\href@noop {} {\emph {\bibinfo {title} {Optical {Waves} in
  {Layered} {Media}}}}\ (\bibinfo  {publisher} {Wiley},\ \bibinfo {year}
  {2005})\BibitemShut {NoStop}%
\bibitem [{\citenamefont {Haus}(1983)}]{haus}%
  \BibitemOpen
  \bibfield  {author} {\bibinfo {author} {\bibfnamefont {H.~A.}\ \bibnamefont
  {Haus}},\ }\href@noop {} {\emph {\bibinfo {title} {Waves and {Fields} in
  {Optoelectronics}}}}\ (\bibinfo  {publisher} {Prentice Hall},\ \bibinfo
  {address} {Englewood Cliffs, NJ},\ \bibinfo {year} {1983})\BibitemShut
  {NoStop}%
\bibitem [{\citenamefont {Kato}(1995)}]{kato}%
  \BibitemOpen
  \bibfield  {author} {\bibinfo {author} {\bibfnamefont {T.}~\bibnamefont
  {Kato}},\ }\href@noop {} {\emph {\bibinfo {title} {Perturbation {Theory} for
  {Linear} {Operators}}}},\ \bibinfo {edition} {2nd}\ ed.\ (\bibinfo
  {publisher} {Springer},\ \bibinfo {address} {Berlin},\ \bibinfo {year}
  {1995})\BibitemShut {NoStop}%
\bibitem [{\citenamefont {Heiss}(2004)}]{heiss_exceptional_2004}%
  \BibitemOpen
  \bibfield  {author} {\bibinfo {author} {\bibfnamefont {W.~D.}\ \bibnamefont
  {Heiss}},\ }\href
  {http://iopscience.iop.org/article/10.1088/0305-4470/37/6/034/meta}
  {\bibfield  {journal} {\bibinfo  {journal} {J. Phys. A: Math. Gen.}\ }\textbf
  {\bibinfo {volume} {37}},\ \bibinfo {pages} {2455} (\bibinfo {year}
  {2004})}\BibitemShut {NoStop}%
\bibitem [{\citenamefont {Mailybaev}\ and\ \citenamefont
  {Seyranian}(2004)}]{mailybaev}%
  \BibitemOpen
  \bibfield  {author} {\bibinfo {author} {\bibfnamefont {A.~A.}\ \bibnamefont
  {Mailybaev}}\ and\ \bibinfo {author} {\bibfnamefont {A.~P.}\ \bibnamefont
  {Seyranian}},\ }\href@noop {} {\emph {\bibinfo {title} {Multiparameter
  {Stability} {Theory} with {Mec}}}}\ (\bibinfo  {publisher} {World Scientific
  Publishing Company},\ \bibinfo {address} {Singapore; River Edge, NJ},\
  \bibinfo {year} {2004})\BibitemShut {NoStop}%
\bibitem [{\citenamefont {Pick}\ \emph {et~al.}(2016)\citenamefont {Pick},
  \citenamefont {Zhen}, \citenamefont {Miller}, \citenamefont {Hsu},
  \citenamefont {Hernandez}, \citenamefont {Rodriguez}, \citenamefont
  {Soljacic},\ and\ \citenamefont {Johnson}}]{pick_general_2016}%
  \BibitemOpen
  \bibfield  {author} {\bibinfo {author} {\bibfnamefont {A.}~\bibnamefont
  {Pick}}, \bibinfo {author} {\bibfnamefont {B.}~\bibnamefont {Zhen}}, \bibinfo
  {author} {\bibfnamefont {O.~D.}\ \bibnamefont {Miller}}, \bibinfo {author}
  {\bibfnamefont {C.~W.}\ \bibnamefont {Hsu}}, \bibinfo {author} {\bibfnamefont
  {F.}~\bibnamefont {Hernandez}}, \bibinfo {author} {\bibfnamefont {A.~W.}\
  \bibnamefont {Rodriguez}}, \bibinfo {author} {\bibfnamefont {M.}~\bibnamefont
  {Soljacic}}, \ and\ \bibinfo {author} {\bibfnamefont {S.~G.}\ \bibnamefont
  {Johnson}},\ }\href {http://arxiv.org/abs/1604.06478} {\bibfield  {journal}
  {\bibinfo  {journal} {arXiv preprint arXiv:1604.06478}\ } (\bibinfo {year}
  {2016})}\BibitemShut {NoStop}%
\bibitem [{SI()}]{SI}%
  \BibitemOpen
  \href@noop {} {}\bibinfo {note} {See Supplementary Information for a
  derivation of the polarization dynamics when $|\tau| = 1$, a derivation of
  the conserved quantities in the scattering matrix of complex birefringent
  materials, a discussion of the polarization dynamics in a lossy material, and
  the effects of off-axis propagation.}\BibitemShut {Stop}%
\bibitem [{\citenamefont {Ge}\ \emph {et~al.}(2012)\citenamefont {Ge},
  \citenamefont {Chong},\ and\ \citenamefont {Stone}}]{ge_conservation_2012}%
  \BibitemOpen
  \bibfield  {author} {\bibinfo {author} {\bibfnamefont {L.}~\bibnamefont
  {Ge}}, \bibinfo {author} {\bibfnamefont {Y.~D.}\ \bibnamefont {Chong}}, \
  and\ \bibinfo {author} {\bibfnamefont {A.~D.}\ \bibnamefont {Stone}},\ }\href
  {\doibase 10.1103/PhysRevA.85.023802} {\bibfield  {journal} {\bibinfo
  {journal} {Phys. Rev. A}\ }\textbf {\bibinfo {volume} {85}},\ \bibinfo
  {pages} {023802} (\bibinfo {year} {2012})}\BibitemShut {NoStop}%
\end{thebibliography}

%

\end{document}